\documentstyle[preprint,aps]{revtex}

\begin{document}
\draft
\tighten

\preprint{\vbox{
\hbox{CWRU-P14-1995}
}}

\title{Perturbed Electroweak Strings and Fermion Zero Modes}

\author{Hong Liu and Tanmay Vachaspati}
\address{
Physics Department\\
Case Western Reserve University\\
Cleveland OH 44106-7079.
}

\date{\today}

\maketitle

\begin{abstract}
The effect of perturbations of electroweak strings on
quark zero modes is studied in 2+1 and
3+1 dimensions. As first discovered by Naculich, it is
found that the bosonic perturbations that destabilize the
string give a mass to the zero modes and also lift their
degeneracy. The effect of the zero modes on the stability of
the string is discussed qualitatively and we argue that
the fermionic vacuum instability found by
Naculich should lead to a distortion of the
bosonic string but not be responsible for decay.
\end{abstract}


\narrowtext

\section{Introduction}

Over the last few years there has been considerable interest
in electroweak $Z-$strings \cite{nambu,tv}.
These are flux tube solutions to the
classical equations of motion of the {\it bosonic}
sector of the standard electroweak model. Recently, interest
has turned to the effect of fermions on the string.

The first issue in including fermions is the effect of the string
on the fermion spectrum. While no one has found the full spectrum,
it is well known that the mass spectrum contains zero energy
solutions (zero modes) for all massive fermions in the
standard model \cite{mewp}.
The zero modes lead to fermionic superconductivity
of the $Z-$string.

The second issue in this research program and also one that we
address in this paper, is to study the backreaction
of fermions on the $Z-$string. Our analysis follows
the recent work of Naculich \cite{steve} who has discussed
the effects of fermions on the stability of an infinite $Z-$string.
The strategy that is adopted is to perturb the string background
and study the effect that this has on the fermion spectrum
and on the fermion ground state energy. This energy contribution
is then included in the total energy and tells us if the fermionic
vacuum contributions to the energy are (i) first or second order in
the perturbations, and, (ii) if they help to stabilize (destabilize)
the string.

In this paper, we consider the back reaction of lepton and
quark zero modes on $Z-$strings and recover some of Naculich's
results. Our techniques and those used in Ref. \cite{steve}
differ in two respects. The first difference is in our
computation of the shift in the fermion energy levels due to
perturbations in the bosonic fields. Naculich employs the
standard result of second order perturbation theory in this
computation in which a sum over a complete set of states needs
to be done. The use of the standard result does not seem
appropriate because the complete spectrum of the fermion modes in the
string background is not known - only the zero modes are known.
So we do not use the standard result and solve the perturbed equations
of motion directly. A second (minor)
difference in our techniques is that we have included all
possible perturbations and not just the perturbation in the
Higgs field.

The result that the fermionic vacuum destabilizes the $Z-$string
in \cite{steve} is due to an infrared logarithmic divergence arising
from low momentum modes propagating along the string. In the
2+1 dimensional case, such modes are absent and it is of some
interest to consider the effect of the fermionic vacuum
(and positive energy fermions) on the string. With this in
mind, we first consider fermions on perturbed strings in 2+1
dimensions and subsequently consider the 3+1 case.
But we find that it is not possible to evaluate the backreaction
of fermions on the string because the full spectrum of
fermion modes is not known. Only the zero mode is known,
and, in 2+1 dimensions, this is only one mode
from an infinity of modes.

While we are not able to evaluate the backreaction quantitatively,
we give an argument why the backreaction of fermions should
lead to a distorted string solution but not be responsible for
string decay. This follows once we realize that the calculations
that have been done for the electroweak string can also be
carried out for low energy {\it topological} strings that can
terminate in superheavy magnetic monopoles. Once
again, the fermionic vacuum will destabilize the topological
string. But, in this case, the topology in the model assures
us that the fermionic vacuum will not cause
the string to decay. Hence, a distorted string solution must
exist. The observation in \cite{kepharttv} that the electroweak
string can arise for topological reasons if the electroweak
symmetry breaking occurred in stages suggests that the
distorted string solution will also exist in the electroweak
model.

We can repeat Naculich's arguments and consider
the effect of fermions on perturbed topological
strings. This, once again, for very general reasons, leads
to an ``instability''. But now the topology indicates that the
fermions will only distort the string and not be responsible
for string decay.

While the present analysis deals with the $Z-$string, we would
like to mention that fermionic effects on analogous vortices
present in $^3$He \cite{voloviktv} have also been
studied. Makhlin and Volovik \cite{makhlinvolovik} find that
fermions distort the structure of the vortices. In future,
we can expect that this distortion might even be observed in the
laboratory \cite{privcommvolovik}.

In Section II, we review past results and set up our
notation. In Section III, we
outline the calculation that ideally needs to be carried
out to determine the back-reaction of fermions on strings.
In Section IV we find the effects of the perturbations on
strings on the fermion zero modes in 2+1 dimensions. Section
V extends this calculation to 3+1 dimensions and we discuss
the fermion backreaction in Section VI.
The Appendix contains a simple explicit proof that fermion zero
modes exist on perturbed Nielsen-Olesen strings in the
Abelian-Higgs model. Though this
follows from the application of an index theorem \cite{ewetal},
our proof is an alternate way of seeing that the zero modes
exist for topological rather than dynamical reasons.

\section{Review of $Z-$string and Fermion Zero Modes}

We consider the standard electroweak model,
\begin{eqnarray*}
{\cal L} ={\cal L}_{Boson} +  {\cal L}_{Fermion}
\end{eqnarray*}
\begin{eqnarray*}
{\cal L}_{Boson}  =  -\frac{1}{4}W_{\mu\nu}^{a}W^{a\mu\nu}
                   -\frac{1}{4}F_{\mu\nu}F^{\mu\nu}
                   +|D_{\mu}\Phi|^{2}
                   -\lambda \left ( \Phi^{\dagger}\Phi -
                    {{\eta^{2}} \over {2}} \right )^{2}
\end{eqnarray*}
where
\[
\Phi = \left(
\begin{array}{c} \phi_{1} \\
                 \phi_{0}
\end{array}
\right),\;\;\;\;
D_{\mu}\Phi  =  (\partial _{\mu}
              - \frac{i}{2}g\tau^{a}W_{\mu}^{a}
              - \frac{i}{2} g' B_{\mu} )\Phi
\]
and $W_{\mu\nu}^{a}$ and $F_{\mu\nu}$ are the corresponding field strength
tensors for $W_{\mu}^{a}$ and $B_{\mu}$. Also define,
\[
\alpha \equiv \sqrt{g^{2} + g'^{2}},\;\;\;\;
\tan\theta_{w} \equiv \frac{g'}{g},\;\;\;\;
e \equiv g\sin\theta_{w} \ .
\]
\[
Z_{\mu}  \equiv  \cos\theta_{w}W_{\mu}^{3}
            - \sin\theta_{w}B_{\mu},\;\;\;\;
A_{\mu}  \equiv  \sin\theta_{w}W_{\mu}^{3}
            + \cos\theta_{w}B_{\mu}
\]
where, as is conventional, the vacuum is
chosen to be $\Phi_{vac}^T = (0, 1)$.

The above model admits classical vortex solutions: $Z-$strings
and $W-$strings. Unlike vortices in the Abelian Higgs model,
these ``embedded''
strings are not topological and they exist only
for energetic rather than topological reasons. Their stability
relies sensitively on the parameters of the theory. When the
analysis is restricted to
the bosonic sector, it has been shown that $Z$ strings are
only stable for small Higgs mass and $\theta_{w}$ close to
$\pi /2$ \cite{mjlptv,mgmh} while $W-$string are always
unstable \cite{mbtvmb}.
A unit winding $Z$-string along the $z$ axis in cylindrical
coordinates is a solution of the form
\[
\Phi = \frac{\eta}{\sqrt{2}}f(r)e^{i\theta}
\left(
\begin{array}{c}  0     \\
                  1
\end{array}
\right),\;\;\;\;
Z_{\theta} = -\frac{2}{\alpha}\frac{v(r)}{r}
\]
with all other components and fields vanishing,
where $f(r)$ and  $v(r)$ satisfy:
\begin{eqnarray}
f'' + \frac{f'}{r}-\frac{f}{r^{2}}(1-v)^{2}
   + \lambda\eta^{2}(1-f^{2})f = 0 \\
v'' - \frac{v'}{r} + \frac{\alpha^{2}\eta^{2}}{4}f^{2}(1-v) = 0
\end{eqnarray}
\[
f(0) = v(0) = 0, \;\;\;\; f(\infty) = v(\infty) = 1
\]

Now we include the fermion sector of the theory. Without loss
of generality, we consider  only quarks in one generation.
\begin{eqnarray*}
{\cal L}_{Quark} = & \bar{\Psi_{L}} i \not\!\!D^{L} \Psi_{L}
                   + \bar{u_{R}} i \not\!\!D^{R} u_{R}
                   + \bar{d_{R}} i \not\!\!D^{R} d_{R}\\
                 &  -  G_{u}(\,\bar{\Psi_{L}}\tilde{\Phi}u_{R}
                     +\bar{u_{R}}\tilde{\Phi}^{\dagger}\Psi_{L}\,)\\
                  & - G_{d}(\,\bar{\Psi_{L}} \Phi d_{R}
                            +\bar{d_{R}}\Phi^{\dagger}\Psi_{L}\,)
\end{eqnarray*}
where
\[
\Psi_{L} = \left(
\begin{array}{c} u_{L} \\
                 d_{L}
\end{array}
\right),\;\;\;\;
\tilde{\Phi} = i\tau_{2}\Phi^{*}=\left(
\begin{array}{c} \phi_{0}^{*} \\
                -\phi_{1}^{*}
\end{array}
\right)
\]
\begin{eqnarray*}
D_{\mu}^{L}\Psi_{L} & = & (\partial _{\mu}
                  - \frac{i}{2}g\tau^{a}W_{\mu}^{a}
                  - \frac{i}{2} g' YB_{\mu} )\Psi_{L} \\
D_{\mu}^{R}\Psi_{R} & = & (\partial _{\mu}
    - \frac{i}{2} g' Y_{R}B_{\mu} )\Psi_{R}
\end{eqnarray*}
In above $\Psi_{R}$ stands for  $u_{R}$ or $d_{R}$ and $Y_{R}$ for their
corresponding hypercharges.

The Dirac equations for the quark fields in the
background of a straight $Z-$string( 2+1 dimension)  are:
\begin{eqnarray}
i \partial_{t} \left(
\begin{array}{c} u\\
                 d
\end{array}
\right) = H_{0} \left(
\begin{array}{c} u\\
                 d
\end{array}
\right) =  \left(
\begin{array}{cc} H_{uu} & 0 \\
                   0     & H_{dd}
\end{array}
\right)  \left(
\begin{array}{c} u\\
                 d
\end{array}
\right)   \label{eq:Dirac}
\end{eqnarray}
with
\begin{eqnarray}
 u  =  \left(
\begin{array}{c} u_{L} \\
                 u_{R}
\end{array}
\right),\;\;\;\;\;\;
  d  = \left(
\begin{array}{c} d_{L} \\
                 d_{R}
\end{array}
\right),\;\;\;\;\;\;
H_{0} = \left(
\begin{array}{cc} H_{uu}  & 0 \\
                      0   & H_{dd}
\end{array}
\right)  \label{eq:zero1}
\end{eqnarray}
\begin{eqnarray}
H_{uu} = \gamma_{0} \left(
\begin{array}{cc}
    -i\not\!d_{u}     & G_{u}\phi_{0}^{*} \\
    G_{u}\phi_{0} &  -i\not\!d_{u}
\end{array}
\right),\;\;\;\;\;\;
H_{dd}  =  \gamma_{0} \left(
\begin{array}{cc}
    -i\not\!d_{d}         & G_{d}\phi_{0} \\
    G_{d}\phi_{0}^{*} &  -i\not\!d_{d}
\end{array}
\right)         \label{eq:zero2}
\end{eqnarray}
\begin{equation}
\not\!d_{u,d} =
\gamma^i (\partial _{i} - i\alpha Z_{i} Q_{z}^{u,d}) \ , \ \ \ \
i=1,2
\label{eq:zero3}
\end{equation}
where $Q_{z}$ are the $Z$ charge matrices
for $u$ and $d$, namely,
\begin{eqnarray*}
Q_{z}^{u,d} & = & \left(
\begin{array}{cc} q^{u,d}_{L} &  0  \\
                     0       &  q^{u,d}_{R}
\end{array}
\right)
       =  \frac{1}{2} \left(
\begin{array}{cc} \pm 1-(y\pm1)\sin^{2}\theta_{w} &  0  \\
                       0      & -(y\pm1)\sin^{2}\theta_{w}
\end{array}
\right)
\end{eqnarray*}
with $y=Y(\Psi^{L})$, the hypercharge of $u_{L}$ and $d_{L}$.
Note that the above equations for $u$ and $d$  are not coupled.

Equation (\ref{eq:Dirac}) has zero energy solutions (``zero modes'')
that are independent of ($z,t$) and are normalizable in the $xy$ plane.
Earlier analyses of zero modes on topological
strings using index theorems \cite{ewetal} find
precisely one zero mode per massive fermion on a string with unit
winding number. This result
holds in our case too and so there exists
precisely one zero mode on a string with winding number 1
for each of $u$ and $d$. Furthermore, the states are
eigenstates of the operator $\gamma^{0}\gamma^{z}$.
As $u$ couples to $\phi_0$ (a vortex) while $d$ couples
to $\phi_0^*$ (an antivortex), we have:
\begin{equation}
    \gamma^{0}\gamma^{z}u_{0} = u_{0},\,\,\,\,\,\,
    \gamma^{0}\gamma^{z}d_{0} = - d_{0}    \label{eq:zero}
\end{equation}
where,
$u_{0}$ and $d_{0}$ denote the zero modes for the $u$ and $d$ fields.
In the context of $3+1$ dimension, the zero energy solution
above generates a whole family of
solutions of  Dirac equations of the form:
\begin{eqnarray}
\Psi_{+} = \beta_{+}(t,z)u_{0}(x,y),\,\,\,\,\,\,
\Psi_{-} = \beta_{-}(t,z)d_{0}(x,y) \label{eq:modes}
\end{eqnarray}
with
\[
(\partial_{t} \pm \partial_{z}) \beta_{\pm}(t,z) = 0
\]
We may think of these solutions as massless chiral fermions
trapped on the string moving at the speed of light: up
quarks move along the string in one direction and we will refer to
them as ``right-movers'' and down quarks move along the string
in the opposite direction and will be called ``left-movers''.
%
%
\section{Outline and Scope of Calculation}

The mutual interaction of fermions and solitons has been
studied over the last few decades and a number of excellent
reviews exist (for example, see Chapter 9 of \cite{rajaraman}).
In this section, we would like to summarize a standard technique
for studying the interaction of fermions with solitons.
Our motivation for this summary is so that we can indicate
the difficulties likely to be encountered in doing the
full calculation and hence indicate the scope of the present
calculation.

The basic idea is that, for a renormalizable theory, Fermi fields
always enter bilinearly in the Lagrangian and can be integrated out
of the functional integral yielding an effective action for the
bosonic fields. If the fermionic part of the action is written
in terms of an operator $K[\Phi , A_\mu ]$ as:
$$
\int d^4 x i {\Psi^{\dag}} K[\Phi , A_\mu ] \Psi
$$
then the effective action for time-independent bosonic fields is:
\[
S_{eff}[\Phi, A_{\mu}] =S_{B}[\Phi, A_{\mu}]
          -i \ln DetK[\Phi, A_{\mu}] \ .
\]
Thus the contributions of the Fermi fields have been entirely
absorbed in the second term and the semiclassical bosonic
solution with fermion back reaction can be obtained by
extremizing $S_{eff}$:
\[
\delta S_{eff}[\Phi, A_{\mu}] =\delta\{S_{B}
-i \ln DetK[\Phi, A_{\mu}]\} = 0
\]
The effects of the Fermi fields become more
transparent  when $DetK[\Phi, A_{\mu}]$ is expressed formally as:
\[
 DetK[\Phi, A_{\mu}] = \sum_{\{n_{r}\}}C(\{n_{r}\})
 \exp[\;-iT(\sum_{r}(-\epsilon_{r} + n_{r}\epsilon_{r}))]
\]
where $\epsilon_{r} = \epsilon_{r}[\Phi, A_{\mu}]$ are the
eigenvalues of the  one particle Dirac Hamiltonion in the
presence of $\Phi$ and $A_{\mu}$, $n_{r}$ are the occupation
numbers for exited states and $C(\{n_{r}\})$ are the combinatoric
degeneracy factors. We recognize that the ground state (with all
$n_{r}=0$) energy $ E_{0}=- \sum_{r}\epsilon_{r}$ is just
the familiar filled ``Dirac sea''.

In practice, it is almost always impossible to implement the
above scheme because  ${\rm ln}DetK[\Phi, A_{\mu}]$ in $S_{eff}$ is,
in general, a non-polynomial non-local functional and its  explicit
form is unobtainable. Instead, to order $\hbar$, it is sufficient
to evaluate the ground state ({\it gs}) energy of the classical
bosonic configuration plus fermionic vacuum as:
\begin{equation}
E_{gs} = E_{0} + \left ( E_b +
           \sum_r (-\epsilon_r ) - E_{ct} \right )
\label{egs}
\end{equation}
where, $E_0$ is the energy of the classical bosonic configuration,
$E_b$ is the contribution due to bosonic fluctuations,
the sum is over all filled energy levels (Dirac sea) with $\epsilon_r$
being the energy eigenvalue of the $r^{th}$ level in the
background of the classical bosonic configuration and
$E_{ct}$ is the energy contribution coming from the counterterms
that are necessary to cancel off the divergences.

In our case, therefore, we would like to calculate all the
fermionic energy levels in the background of the $Z-$string
and the energy contribution from the counterterms. That will
yield $E_{gs}-E_b$. Then we would like to do the same calculation
with a slightly perturbed $Z-$string; this will yield
$E_{gs}-E_b +\delta E$. If $\delta E$ is negative, the fermionic
vacuum destabilizes the string; if it is positive, we say that
the fermionic vacuum tends to stabilize the string.

This simplified calculation is also beyond our reach for we
do not know all the energy levels $\epsilon_r$. We only know
a part of the spectrum - the zero modes on the string. If we
{\it assume} that the dominant contribution to the term in parenthesis
in (\ref{egs}) comes from the zero modes we might be able to say
something about the effects of fermions on the string
configuration\footnote{Intuitively it seems reasonable that a
fermion scattering state should give a smaller contribution
to $\delta E$ than a bound state.
This is because the scattering
states live outside the string core and have little overlap
with perturbations of the string core. But it is not known
if the zero modes are the only bound states and if the
summed effect of all the scattering states is smaller
than the summed effect of the bound states since there
are many more scattering states than there are bound states.}.
Indeed this calculation was performed in
\cite{steve} with the conclusion that $\delta E$ due to the
fermionic vacuum is always negative. We will outline this
calculation in Sec. VI and argue that this probably leads
to a distortion of the $Z-$string.



\section{Perturbed Zero Modes in 2+1 Dimension}

In the case of topological strings, the existence of fermionic
zero modes follows from an index theorem. Then it is clear that
the zero modes will continue to exist even if the topological
string solution is perturbed. For the same reason, since the
$Z-$string is a topological string
when we restrict ourselves to the $U(1)$ sector defined by
the $Z$ gauge field, the zero modes continue to exist when only
the $\phi_{0}$, $Z$ and $A$ fields are perturbed.
(A simple explicit
argument to see this without recourse to the index theorem is
described in the Appendix.) However, in the full model, perturbations
involving $W^{\pm}$ and $\phi^{+}$ fields take one out of the
$U(1)$ sector and we may expect non-trivial consequences for
the zero modes.

According to \cite{mjlptv}, the physical modes that destabilize
the $Z-$string involve combinations
of $W_{\pm}$ and $\phi_{1}$ modes in the $xy$ plane.
Therefore,  we will restrict our discussions to perturbation of
this type, {\it i.e.} we assume
\begin{eqnarray}
\phi_{1} = \epsilon \phi_{1}(x,y),\;\;\;\;\;
W^{\mu}  = \epsilon (0,W^{1}(x,y),W^{2}(x,y),0) \label{eq:per}
\end{eqnarray}
where $\phi$, $W^{1}$, $W^{2}$ are arbitrary complex functions of $x$ and
$y$ and $\epsilon$ is a perturbation parameter. Note that we
have not fixed the gauge as this is not necessary for our
purposes. (For an alternate approach to the stability analysis
see \cite{mgmh}.)

In  (2+1) dimension, with  $\Psi = \psi(x,y)e^{-iEt}$,
the Dirac equations for quarks\footnote{The leptons can be similarly
considered.} in the background of the
perturbed string become energy eigenvalue equations:
\begin{eqnarray}
H \,\Psi = E\,\Psi
\label{eq:1}
\end{eqnarray}
with
\begin{eqnarray}
\Psi  =  \left(
\begin{array}{c} u\\
                 d
\end{array}
\right),\;\;\;\;\;\;
  H  =  H_{0} + \epsilon H_{1},\;\;\;\;\;\;
 H_{1}  =  \left(
\begin{array}{cc}    0    & H_{ud}   \\
                  H_{du}  &  0
\end{array}
\right)  \label{eq:H1}
\end{eqnarray}
\[
H_{ud} =  H_{du}^{\dagger} =  \gamma_{0} \left(
\begin{array}{cc}
          -\frac{g}{2}\not\!W^{+}         & G_{d}\phi_{1} \\
     - G_{u}\phi_{1} &  0
\end{array}
\right) \;\;\;\;\;\;
W^{\pm} = W^{1}\mp iW^{2} \;\;\;\;\;\;\;
\not\!W^{+} = \gamma^{\mu}W^{+}_{\mu}
\]
where $u,d,H_{0}$ have the same meaning as in (\ref{eq:zero1}),
(\ref{eq:zero2}), (\ref{eq:zero3}) and $\phi_{1}$, $W^{\pm}$ are
given by (\ref{eq:per}). Notice that the perturbations in $W$ and
$\phi_{1}$ introduce couplings between the $u$ and $d$ fields.

A crucial property of the Hamiltonian, and one which we
will repeatedly use, is:
\begin{equation}
  \{\gamma^{0}\gamma^{z}\, , \, H\}  =  0  \label{eq:symmetry}
\end{equation}
where \{\} stands for the anticommutator.
The origin  of this property can be traced to the two dimensional
character of the field configuration and its existence
can be expected in any generic (2+1) dimensional problem.
In our case, (\ref{eq:symmetry}) is true both before and after the
string is perturbed. From (\ref{eq:symmetry}), the transformation:
$\Psi\,\rightarrow \gamma^{0}\gamma^{z}\Psi$
(called  ``particle conjugation'' in \cite{rjcr})
takes a solution with energy $E$ to one with energy $-E$,
{\it i.e.} if
\begin{equation}
H \Psi  = E \Psi  \label{eq:key1}
\end{equation}
then
\begin{equation}
H(\gamma^{0}\gamma^{z}\Psi) =   -E(\gamma^{0}\gamma^{z}\Psi) \ .
 \label{eq:key2}
\end{equation}
Also, since $H$ is hermitian, $\Psi$ and
$\gamma^{0}\gamma^{z}\Psi$ are orthogonal to each
other except when $E=0$.

Now we investigate the behaviour of the zero modes in the context of
perturbation theory. When $\epsilon=0$, we recover (\ref{eq:Dirac}),
where the  equations for $u$ and $d$ are uncoupled and there are
two {\it degenerate} zero modes
\[
\left (
\begin{array}{c} u_{0} \\
                 0
\end{array}
\right ),\;\;\;\;\;
\left (
\begin{array}{c} 0 \\
                 d_{0}
\end{array}
\right ) \ .
\]

When the string is perturbed,
the energy eigenstates are linear combinations of
$(u_0, 0)^T$ and $(0,d_0 )^T$ and so we write them as:
\[
\Psi_{0} = a \left (
             \begin{array}{c} u_{0} \\
                                0
             \end{array}
             \right )
         + b \left (
             \begin{array}{c} 0 \\
                             d_{0}
             \end{array}
             \right )
\]
where, $a$ and $b$ are constants that need to be determined.
The equation for $\Psi_0$ is
\begin{eqnarray}
H\;\Psi_{0} = (H_{0} + \epsilon H_{1})\;\Psi_{0} =
   (E_{0} + \epsilon E_{1}) \;\Psi_{0} \label{eq:zerosecond}
\end{eqnarray}
and is straightforward to solve. This gives:
\begin{equation}
       E_{1} = \pm | m_{1} |,\;\; \frac{a}{b} =
        \pm  \frac{{m_{1}}}{|m_{1}|} \equiv \pm e^{i\beta}
\label{eone}
\end{equation}
where $m_{1}$ is defined by
\begin{eqnarray}
 {m_{1}} \equiv \; <u_{0}|H_{ud}|d_{0}>\;
       = \int\!d^{2}x \,u_{0}^{\dagger}H_{ud}d_{0} \ .
\label{eq:m1}
\end{eqnarray}
The magnitude of $a$ and $b$ is not fixed by perturbation
theory but can be fixed by imposing a suitable normalization
condition.

Eq. (\ref{eone}) shows that the perturbations on the string
lift the two-fold degeneracy of zero modes yielding two
states with opposite energies. In the limit that the perturbation
is turned off ($\epsilon \rightarrow 0$), the two eigenstates with
their $H_1$ eigenvalues are:
\begin{equation}
\Psi_{0}  =  \left (
\begin{array}{c} u_{0} \\
                 e^{-i\beta}d_{0}
\end{array}
\right ), \ \ \ \
E_{1} =  + |m_{1}|
\label{psi0one}
\end{equation}
\begin{equation}
\tilde{\Psi}_{0}  = \left (
\begin{array}{c} u_{0} \\
                -e^{-i\beta}d_{0}
\end{array}
\right ), \ \ \ \
\tilde{E_{1}} = - |m_{1}|
\label{psi0two}
\end{equation}
Note  that $\tilde{\Psi}_{0}$ and $\Psi_{0}$ satisfy
\begin{equation}
\tilde{\Psi}_{0}  =  \gamma^{0}\gamma^{z}\Psi_{0} \ ,\ \ \ \
\tilde{E_{1}} = -E_{1}
\label{etominuse}
\end{equation}
a consequence of (\ref{eq:key1}) and (\ref{eq:key2}).
Actually, (\ref{eq:key1}) and (\ref{eq:key2}) require
that the relation (\ref{etominuse})
holds to all orders in perturbation theory, {\it i.e.} if we
denote  the set of corrections to the wave function and
energy of $\Psi_{0}$ as  $\{\Psi_{n},\, E_{n},\,n=1,2,\ldots\}$
and that of  $\tilde{\Psi}_{0}$ as
$\{\tilde{\Psi}_{n},\, \tilde{E_{n}},\, n=1,2,\ldots\}$,
then
$$
{\tilde \Psi}_n = \gamma^0 \gamma^n \Psi_n \ ,
\ \ \
{\tilde E}_n = - E_n
$$
for all $n$.

In analyzing the effect of the fermions on the stability
of the string, it is necessary to work up to second order in
perturbation theory. So we now need to find $E_2$.

First note that the conventional result for the second order
change in energy of a state $|n>$ is written as:
\begin{equation}
E_{2}^{(n)} = \sum_{m \neq  n}\frac{|<m|H'|n>|^{2}}
{E^{(m)}_0 -E^{(n)}_0}
\label{perttheory}
\end{equation}
where, $E^{(i)}_0$ denotes the energy of the unperturbed
$i^{th}$ level. This sum requires knowledge of the entire
unperturbed spectrum $E^{(i)}_0$ and hence cannot be used
in our case where we only know one state, namely,
the zero mode. So we need to find $E_2$ by some
other means. The means we adopt is to make crucial use
of the property in (\ref{eq:symmetry}).

Let us denote
the perturbative corrections to the wave function and
energy of the zero mode by $\{\Psi_{n},\, E_{n},\,n=1,2,\ldots\}$
and that of  $\tilde{\Psi}_{0}$ as
$\{\tilde{\Psi}_{n},\, \tilde{E_{n}},\, n=1,2,\ldots\}$.
$E_{2}$ (similarly $\tilde{E_{2}}$) can be obtained by:
\[
  E_{2} = <\Psi_{0}|H_{1}|\Psi_{1}> \
\]
but we first have to find $\Psi_1$.
If we write
\begin{eqnarray}
\Psi_{1}  =  \left(
\begin{array}{c} u_{1}\\
                 d_{1}
\end{array}
\right) \label{eq:a1}
\end{eqnarray}
then $u_1$ and $d_1$ satisfy:
\begin{eqnarray}
 H_{uu}u_{1} & =  & b {m}_{1}u_{0} -b H_{ud}d_{0}
       \label{eq:p1} \\
 H_{dd}d_{1} & = & a {m}_{1}^{*}d_{0} -a H_{du}u_{0}
       \label{eq:p2}
\end{eqnarray}
and,
\begin{eqnarray}
E_2 = a^* < u_0 | H_{ud} | d_1 > + b^* < d_0 | H_{du} | u_1 >
      \label{eq:E2}
\end{eqnarray}
Equations (\ref{eq:p1}) and (\ref{eq:p2}) are impossible to solve
(even numerically) since they involve the unspecified perturbations
$W^{\mu}$ and $\phi^+$ entering via ${m}_{1}$, $H_{ud}$
and $H_{du}$. Yet, as we show below, it is still possible to
find $E_2$ (and $\tilde E _2$) by using the property
in (\ref{eq:symmetry}).

Let
\begin{eqnarray}
u_{1} = u_{1}^{+} + u_{1}^{-},\;\;\;\;
d_{1} = d_{1}^{+} + d_{1}^{-} \label{eq:ud}
\end{eqnarray}
where
\begin{eqnarray}
\gamma^{0}\gamma^{z}u_{1}^{\pm} = \pm u_{1}^{\pm},\;\;\;\;
\gamma^{0}\gamma^{z}d_{1}^{\pm} = \pm d_{1}^{\pm}
\label{eq:ud1}
\end{eqnarray}
Then, using the  property in (\ref{eq:symmetry}) together with
$(\gamma^{0}) ^{\dag} = \gamma^{0}$ and $(\gamma^{z}) ^{\dag}
= -\gamma^{z}$,
we get,
\begin{equation}
< u_{0} | H_{ud} | d_{1}^{+} > = - < u_{0} |
(\gamma^{0} \gamma^{z} )^{\dag} H_{ud}
(\gamma^{0} \gamma^{z} )|d_{1}^{+} >
\end{equation}
Now we use eqns. (\ref{eq:zero}) and (\ref{eq:ud1})
which give:
\begin{equation}
< u_{0} | H_{ud} | d_{1}^{+} > = 0 \label{eq:E31}
\end{equation}
Similarly, we find
\begin{equation}
< d_{0} | H_{du} | u_{1}^{-} > = 0 \label{eq:E32} .
\end{equation}
Then (\ref{eq:E2}), (\ref{eq:E31}) and (\ref{eq:E32}) give
\begin{eqnarray}
 E_{2} = a^{*}<u_{0}|H_{ud}|d_{1}^{-}>
       + b^{*}<d_{0}|H_{du}|u_{1}^{+}> \label{eq:E4}
\end{eqnarray}
Note that $u_{1}^{-}$ and $d_{1}^{+}$ do not contribute to $E_{2}$
(or to $\tilde{E_{2}}$).
In terms of $u_{1}^{\pm}$,$d_{1}^{\pm}$,
equations (\ref{eq:p1}) and (\ref{eq:p2}) reduce to:
\begin{eqnarray*}
H_{uu} u_{1}^{+} & = & 0, \;\;
H_{uu}u_{1}^{-}  =  b {m}_{1}u_{0} -b H_{ud}d_{0} \\
H_{dd} d_{1}^{-} & = & 0, \;\;
H_{dd} d_{1}^{+}  =  a {m}_{1}^{*}d_{0} -a H_{du}u_{0}
\end{eqnarray*}
We see that the equations for $u_{1}^{+}$ and $d_{1}^{-}$
are just the zero modes equations of
$H_{0}$ (see (\ref{eq:Dirac})) and so the solutions are
$u_{1} ^{+} \propto u_{0}$ and $d_{1} ^{-} \propto d_{0}$.
But since we have fixed the normalization of the
wave-function, we require $ <\Psi_{0}|\Psi_{1}> = 0 $.
Therefore $u_{1}^{+}$ and $d_{1}^{-}$ have to vanish. From
(\ref{eq:E4}) (and similarly for $\tilde{E_{2}}$) we
then get
\[
   E_{2} = \tilde{E_{2}} = 0
\]

%



\section{Perturbed Zero Modes in 3+1 dimension}

In a (3+1) dimensional context, equation (\ref{eq:1})
becomes:
\[
 ( H - i\gamma^{0}\gamma^{z}\partial_{z})\Psi = E \Psi
\]
where $H=H(x,y)$ is defined as before with no dependence on $z$.
In this case, it is straightforward to obtain the perturbed zero
modes moving along the string by Lorentz boosting  the perturbed
solutions in (2+1) dimension that we found in the last section
along the $z$ direction. Considering only the lowest order correction,
we get two sets of levels:
\begin{equation}
E = \sqrt{P^{2} + \epsilon^{2}|m_{1}|^{2}}
\label{root1}
\end{equation}
by boosting $\Psi_{0}$ and
\begin{equation}
{\tilde E} = - \sqrt{P^{2} + \epsilon^{2}|m_{1}|^{2}}
\label{root2}
\end{equation}
by  boosting $\tilde{\Psi}_{0}$, where, $P$ is the momentum in
the $z$ direction. This is
just the energy spectrum for a particle moving freely along the
string with momentum $P$ and mass $\epsilon |m_{1}|$.

Note that
we only need to boost the {\it first} order correction to the
zero mode ($P=0$) in order to get the {\it second} order
correction to the $P \ne 0$ modes. Strictly, this follows
only for $P^2 >> \epsilon^2 |m_1 |^2$. This assumption is
justified
if we assume periodic boundary conditions
in the $z$ direction with period $L$ and assume $\epsilon$ small
enough so that $\epsilon m_1 L  < < 1$. (For an infinite string,
we need to take the limit $L \rightarrow \infty$ as well as
$\epsilon \rightarrow 0$ for perturbation theory to be applicable.
The necessity of taking two limits requires care and we always assume
that they are taken so that $\epsilon |m_1| L$ remains small.)

The above results for $E$ and $\tilde E$
can also be reached by considering perturbations
around the massless modes in (\ref{eq:modes}). Since
\begin{equation}
   [\, H , P_{z}\,\,] =0 \ ,
\label{hpz}
\end{equation}
we can choose $\Psi$ to be an eigenstate of $P_{z}$, {\it i.e.}
\[
 \Psi \rightarrow e^{iPz}\Psi(t,x,y) \ .
\]

Our unperturbed massless states are:
\[
E_{0} = P,\;\;\;\;\;\;  \Psi_{0} = e^{i(-E_0 t + Pz)} \left(
                    \begin{array}{c} u_{0} \\
                                      0
                     \end{array}
                    \right)
\]
\[
E_{0} = -P,\;\;\;\;\;\;  \Psi_{0} = e^{i(-E_0 t + Pz)} \left(
                    \begin{array}{c} 0 \\
                                     d_{0}
                     \end{array}
                    \right)
\]

Now let us look at the perturbations of the above states when the
strings are perturbed. (We will only discuss the case
$E_{0} = P$; identical procedures apply to the case
$E_{0} = -P$.)

Notice that because $P_z$ and $H$ commute
(eq. (\ref{hpz})), we can restrict our attention to sectors
of fixed $P_z$ when doing the perturbation analysis.
In such a sector with $P \ne 0$, the zeroth order wave functions
are not degenerate. Now since
$ H_{1}$ (see (\ref{eq:H1})) is off diagonal, and
the zeroth order wave function with $E_0 = +P$, $\Psi_{0}$,
is proportional to $( u_{0}, 0)^T$,  the first order energy
perturbation $E_{1} = <\Psi_{0}|H_{1}|\Psi_{0}> =  0$.

The energy perturbation will be non-trivial in the second
order calculation but, as in the
(2+1) dimensional case, the standard result of
second order perturbation theory cannot be used here. Once
again this difficulty can  be avoided by the same technique
used in Sec. IV.
Using  the same notation as in the (2+1) dimensional
case (see (\ref{eq:a1})), now the equations for $u_{1}$
and $d_{1}$ become:
\begin{eqnarray*}
 H_{uu}u_{1} & =  & P(1 - \gamma^{0}\gamma^{z})u_{1}\\
 H_{dd}d_{1} & = &  P(1 - \gamma^{0}\gamma^{z})d_{1}
                     - H_{du}u_{0}
\end{eqnarray*}
Decomposing $u_{1}$ and $d_{1}$ as in (\ref{eq:ud}) and
(\ref{eq:ud1}), the above equations become:
\begin{eqnarray}
H_{uu} u_{1}^{-} & = & 0, \;\;
H_{uu}u_{1}^{+}  =  2Pu_{1}^{-} \label{eq:2} \\
H_{dd} d_{1}^{-} & = & 0, \;\;
H_{dd} d_{1}^{+}  =  2Pd_{1}^{-} - H_{du}u_{0}
\label{eq:3}
\end{eqnarray}
Since $H_{uu}$ does not have any zero energy solution
which is also an eigenstate of $\gamma^{0}\gamma^{z}$ with
eigenvalue $-1$, it follows from (\ref{eq:2}) that
$u_{1}^{-} = 0$, and the equation for $u_{1}^{+}$ again
reduces to that of the zero modes. With the imposition of
the normalization condition, we conclude  $u_{1}^{+} = 0 $.

Next let us look at (\ref{eq:3}). The equation
for $d_{1}^{-} $ leads to:
\[
     d_{1}^{-} = N d_{0}
\]
where $N$ is some  constant to be determined.
(Note that the normalization condition $<\Psi_0 | \Psi_1 > =0$
does not impose any constraint on $N$.)
Plugging the above expression for $d_{1}^{-}$
into the equation for $d_{1}^{+}$, taking
the inner product of the equation with $d_{0}$ and
using $H_{dd} d_0 = 0$ (see (\ref{eq:Dirac})), we get:
\[
2PN = <d_{0}|H_{du}|u_{0}>
\,\,\,\, \Rightarrow
N = \frac{{m}_{1}^{*}}{2P}
\,\,\,\, \Rightarrow
d_{1}^{-} = \frac{{m}_{1}^{*}}{2P} d_{0}
\]
where ${m}_{1}$ is defined as before (see (\ref{eq:m1})).
Then, it follows that:
\begin{equation}
E_{2} = <\Psi_{0}|H_{1}|\Psi_{1}>
        = <u_{0}|H_{1}|d_{1}^{-}>
  =  \frac{|m_{1}|^{2}}{2P} \ .
\end{equation}
Therefore our final results are:
\begin{equation}
E = E_{0} + \epsilon E_{1} + \epsilon^2 E_{2}
  = P + \epsilon^2  \frac{|m_{1}|^{2}}{2P}
\label{e1}
\end{equation}
for left-movers, and,
\begin{equation}
E = E_{0} + \epsilon E_{1} + \epsilon^2 E_{2}
  = - P -  \epsilon^2  \frac{|m_{1}|^{2}}{2P}
\label{e2}
\end{equation}
for right-movers.
These expressions agree with (\ref{root1}) and (\ref{root2}) to
second order in $\epsilon^2$ for $P^2 >> \epsilon ^2 |m_1|^2$.
We exhibit these results in Fig. 1.

In \cite{steve}, the results in (\ref{e1}) and (\ref{e2})
were also obtained but by applying the standard formula of
second order perturbation theory (\ref{perttheory}).

\section{Conclusions and Discussion}

We have explored the effects of perturbations on $Z-$strings
on fermion zero modes in 2+1 and 3+1 dimensions.
For the $P=0$ state, we used degenerate
perturbation theory to find the first order correction to
the energy of the zero mode. A difficulty which we encountered
when trying to find the second order corrections to the energy
(for any value of $P$) was
the unavailability of a complete basis of zeroth order wave functions.
The standard formula of second order perturbation theory assumes the
knowledge of such a basis. We circumvented this difficulty
by using symmetry arguments and by extracting the relevant
part of the solution to the first order equations. This then
gave us the second order correction to the energy for all
values of $P$.

Having found the effect that perturbations
on electroweak strings have on fermion zero modes we now
wish to discuss the backreaction of fermions on the string.
This problem has been discussed by Naculich \cite{steve}
and we first summarize his result.

The fermionic ground state consists of the Dirac sea in
which all negative energy levels are filled. Then the
effect of the perturbations can be found by summing
the shifts in all the negative energy levels. From (\ref{e2}),
this leads to a sum of the kind:
\begin{equation}
\Delta E = -{{\epsilon^2} \over 2} |m_1|^2 L
\sum_{n=1} ^N {1 \over n}
\label{div}
\end{equation}
where periodic boundary conditions have been imposed along
the string with period $L$, an ultraviolet cut-off on the
sum has been imposed and the two
degenerate states with $P=0$ have been taken to be both filled
or both empty. The sum in (\ref{div})
is logarithmically divergent.
Furthermore, the shift in the energy is negative
since all the negative energy modes have shifted down
(see Fig. 1). The divergence in this contribution will
be tamed once the contribution of the counterterms (see
eq. (\ref{egs})) is taken into account but the sign remains
negative for an infinite string. This indicates that
the fermionic vacuum destabilizes the bosonic electroweak
string. As discussed in \cite{steve}, this could mean
that the solution for the string - found by solving the
classical equations in the bosonic sector - gets distorted
by the fermionic vacuum or it might imply that there is a
runaway instability and the string decays into the vacuum.
We now argue that this instability is likely to distort
the string but not lead to decay.

The instability discussed above relies on three
features of the system on hand. These are:
(i) the existence of fermionic zero modes on
the unperturbed strings, (ii) the existence of certain
perturbations that give the massless modes a small
mass as in eqs. (\ref{root1}) and (\ref{root2}), and,
(iii) the logarithmic divergence in the sum
over the Dirac sea. As these features are present in a wide
variety of defects, the fermionic vacuum instability
can be expected to apply in very general
situations. In particular, suppose
we have a model in which topological magnetic
monopoles are produced at extremely high energy scales
such as near the Planck scale and then get connected
by strings at a low energy scale such as the electroweak
scale. Then, in the low energy theory, the strings are the
usual {\it topological} strings and we can consider a
fermionic sector such that fermionic
zero modes exist on these strings. This model has
all the features necessary for the fermionic vacuum
instability. By assumption, it has the feature that there
are fermionic zero modes on the strings. This implies that
the fermions are massive
outside the string and massless within. But then, since
the strings can terminate, the distortions of the string
that are present near the terminus (monopole) must give
a mass to the fermions exactly as in (\ref{root1})
and (\ref{root2}). Once this feature is present, the
logaritmic divergence with the negative sign simply
follows from summing over the Dirac sea exactly
as in (\ref{div}). This shows that
even when the monopoles are extremely heavy (Planck
scale) and the strings are very light (electroweak
scale), the fermionic vacuum instability persists.
However, in this case, we know that the only way the string can
decay is by nucleating monopole-antimonopole pairs because
of the topology in the model. Given that the monopole
is very heavy, the nucleation is an
exponentially suppressed quantum tunneling
process. So the instability due to the fermionic
vacuum cannot lead to string decay but can only distort
the string configuration.

The electroweak string can also be viewed as a topological
string that can terminate on monopoles \cite{kepharttv} except
that there is no vast separation of scales as in the
above example. But this separation of scales does not
enter the fermionic vacuum instability and is irrelevant
for present purposes. So we expect that the fermionic
vacuum contributions found in \cite{steve} should
lead to a distortion of the electroweak string but not
be responsible for decay.

Another argument that supports this conjecture
comes from the fact that the
instability does not depend on whether the string is global
or local and so it should also apply in condensed matter
systems where fermionic zero modes exist on global strings.
$^3$He is one such system in which topologically unstable
global strings with fermionic zero modes exist \cite{voloviktv}
and such strings have been observed in the laboratory. It has
been argued \cite{makhlinvolovik} that the fermionic vacuum
should distort the structure of these strings. Perhaps we will
be able to observe the fermionic vacuum distortion of strings
in $^3$He in the near future.

\

\

\noindent{\bf Acknowledgements}

We are grateful to Steve Naculich and Grisha Volovik
for enlightening discussions.

\vfill
\eject

\centerline{\bf Appendix}

\

\

We now give a simple explicit proof that the zero modes on
Nielsen-Olesen vortices in the Abelian Higgs model continue
to exist even when the vortices undergo deformations (which
are not necessarily small). The proof is based on the
property (\ref{eq:symmetry}) of the Hamiltonian.

Consider an eigenstate $\psi$ of the Dirac Hamiltonian
in a deformed string background. Then we write
$$
H = H_0 + \epsilon H_1
$$
where $H_0$ is the Hamiltonian in the undeformed string
background, $\epsilon$ is a parameter (not necessarily
small) and $\epsilon H_1$ is the Hamiltonian due to the
deformations of the string.
Then we have
$$
H \psi = E \psi \ .
$$
Next define:
$$
\psi_0 \equiv \lim_{\epsilon \rightarrow 0}\psi \ .
$$
We will first show that $\psi_0$ cannot be an eigenstate
of the operator $\gamma^0 \gamma^z$ if $E \ne 0$.

Eqns. (\ref{eq:key1}) and (\ref{eq:key2}) and continuity
in the parameter $\epsilon$ give
\[
<\psi_0 |\gamma^{0}\gamma^{z}\psi_0 > =
\lim_{\epsilon \rightarrow 0}<\psi|\gamma^{0}\gamma^{z}\psi> =0 \ .
\]
This says that the states $\psi_0$ and $\gamma^0 \gamma^z \psi_0$
are orthogonal.
But if $\psi_0$ were an eigenstate of $\gamma^0 \gamma^z$
with eigenvalue $c$ we would have,
\[
<\psi_{0}|\gamma^{0}\gamma^{z}\psi_{0}> =
c <\psi_{0}|\psi_{0}> =  c
\]
which contradicts the orthogonality\footnote{Note that $c$ cannot
vanish because $\gamma^0 \gamma^z$ is unitary.}.

In the Abelian Higgs model, as long as left- and right-moving
fermions do not couple directly in the fermionic sector of
the Lagrangian, $\psi_0$ - being either a left-mover or
a right-mover - is necessarily an eigenstate of
$\gamma^0 \gamma^z$ and, hence, $E=0$ for all values of
$\epsilon$. In the electroweak model, the left- and
right-movers (down and up type quarks)
do not couple directly if we restrict ourselves
to the $Z$ sector of the model. And so,
the zero modes will persist when the string is deformed
within the $Z$ sector. However, once we allow perturbations
outside the $Z$ sector, the left- and right-movers do
couple directly via terms such as ${\bar u}_L \phi_1 d_R$ and
we can expect the zero modes to develop a mass. This is
also clear from eqns. (\ref{psi0one}) and (\ref{psi0two}),
where we see that $\psi_0$ is not an eigenstate of
$\gamma^0 \gamma^z$.

\vfill
\eject

\centerline{\bf Figure Caption}

\

\

1. The spectrum of massless quarks on electroweak
strings (straight lines)
and when the strings are perturbed (hyperbolae).


\begin{thebibliography}{999}

\bibitem{nambu} Y. Nambu, {\it Nucl. Phys. B} {\bf 130} (1977) 505.

\

\bibitem{tv} T. Vachaspati,
{\it Phys. Rev. Lett.} {\bf 68} (1992) 1977-1980,
{\it ibid} {\bf 69} (1992) 216E.

\

\bibitem{mewp} M. Earnshaw and W. Perkins,
{\it Phys. Lett. B} {\bf 328} (1994) 337.

\

\bibitem{steve} S. Naculich,
{\it Phys. Rev. Lett.} {\bf 75} (1995) 998-1001; S. Kono and
S. Naculich, {\it hep-~ph/9507350}.

\

\bibitem{kepharttv} T. Kephart and T. Vachaspati,
{\it hep-ph/9503355}.

\

\bibitem{voloviktv} G.E. Volovik and T. Vachaspati,
{\it cond-mat/9510065}.

\

\bibitem{makhlinvolovik} Y.G. Makhlin and G.E. Volovik,
{\it cond-mat/9510075}.

\

\bibitem{privcommvolovik} G.E. Volovik, {\it private communication}.

\

\bibitem{ewetal} E. Weinberg, {\it Phys. Rev. D} {\bf 24} (1981) 2669.

\

\bibitem{mjlptv} M. James, L. Perivolaropoulos and T. Vachaspati,
{\it Nucl. Phys. B} {\bf 395} (1993) 534-546.

\

\bibitem{mgmh} M. Goodband and M. Hindmarsh, {\it hep-ph/9505357}.

\

\bibitem{mbtvmb} M. Barriola, T. Vachaspati and M. Bucher,
{\it Phys. Rev. D} {\bf 50}, (1994) 2819-2825.

\

\bibitem{rajaraman} R. Rajaraman,
{\it Solitons and Instantons} North-Holland (1982).

\

\bibitem{rjcr} R. Jackiw and P. Rossi,
{\it Nucl. Phys. B} {\bf 190} (1980) 681.

\end{thebibliography}
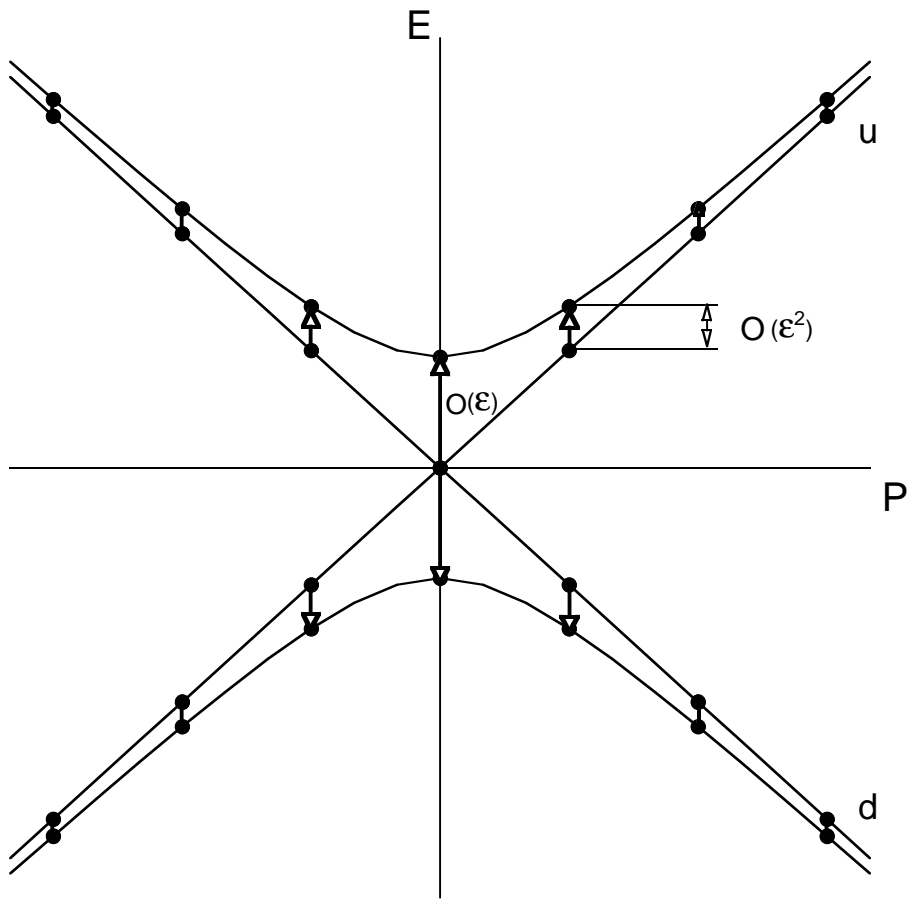
\end{document}